\documentclass[12pt, journal, onecolumn]{IEEEtran}

\usepackage{subfig}
\usepackage{url}
\usepackage{ifthen}
\usepackage{cite}
\usepackage[cmex10]{amsmath} 
\usepackage{bbold}

\usepackage{mathtools}
\usepackage{balance}
\usepackage{amsthm}
\usepackage{multirow}
\usepackage{amsfonts}
\usepackage{graphicx}
\usepackage{algorithm}
\usepackage{url}
\usepackage{lipsum}
\usepackage{booktabs,makecell,multirow}
\usepackage{tikz}
\usetikzlibrary{matrix}
\usepackage{rotating}

\usepackage{latexsym}
\usepackage{color}
\usepackage{graphicx}
\usepackage{threeparttable}
\usepackage{float}

\usepackage{algpseudocode}

\errorcontextlines\maxdimen

\usepackage{bookmark}

\DeclareGraphicsExtensions{.tif,.pdf,.jpeg,.png,.jpg,.bmp,.svg,.eps,.EMF}

\newtheorem{theorem}{Theorem}
\newtheorem{lemma}{Lemma}

\newcommand{\cM}{\mathcal{M}}

\newcommand{\cS}{\mathcal{S}}
\newcommand{\cT}{\mathcal{T}}

\newcommand{\cL}{\mathcal{L}}

\newcommand{\cV}{\mathcal{V}}
\newcommand{\cH}{\mathcal{H}}

\newcommand{\bX}{\mathbf{x}}
\newcommand{\bY}{\mathbf{y}}

\newcommand{\bV}{\mathbf{v}}

\newcommand{\bA}{\mathbf{a}}
\newcommand{\bB}{\mathbf{b}}

\newcommand{\bS}{\mathbf{s}}

\newcommand{\bL}{\mathbf{l}}

\newcommand{\poly}{\mathsf{poly}}

\newcommand{\toInt}{\mathsf{Vec2Int}}

\newcommand{\wt}{\mathsf{wt}}

\newcommand{\Mod}[1]{\ \mathrm{mod}\ #1}

\DeclarePairedDelimiter\ceil{\lceil}{\rceil}


\begin{document}

%
%
%

\title{Improved algorithms for non-adaptive group testing with consecutive positives}

\author{Thach V.~Bui,~\IEEEmembership{Member,~IEEE,}
        Mahdi~Cheraghchi,~\IEEEmembership{Senior Member,~IEEE,}
        An T.H.~Nguyen,~\IEEEmembership{Non-member,~IEEE,}
        and~Thuc D.~Nguyen,~\IEEEmembership{Non-member,~IEEE}
\thanks{Thach V. Bui was supported in part by Vietnam National University Ho Chi Minh City (VNU-HCM) under Grant No.\ NCM2019-18-01. Thuc D. Nguyen was supported in part by Vietnam National University Ho Chi Minh City (VNU-HCM) under Grant No.\ NCM2019-18-01 and by University of Science, VNU-HCM under Grant No.\ CNTT 2021-27. M.\ Cheraghchi's research was partially supported by the National Science Foundation under Grant No.\ CCF-2006455. The material in this paper was presented in part at the 2021 IEEE International Symposium on Information Theory~\cite{bui2021improved}.}
\thanks{Thach V.~Bui is with the Department of Computer Science, National University of Singapore, Singapore 117417, on leave from the Faculty of Information Technology, University of Science, VNU-HCMC, Ho Chi Minh City 72711, Vietnam and Vietnam National University, Ho Chi Minh City 720300, Vietnam (e-mail: bvthach@fit.hcmus.edu.vn).}
\thanks{Mahdi~Cheraghchi is with the Department of EECS, University of Michigan, Ann Arbor, MI 48109, USA (e-mail: mahdich@umich.edu).}
\thanks{An T.H.~Nguyen was with the Faculty of Information Technology, University of Science, VNU-HCMC, Ho Chi Minh City 72711, Vietnam and Vietnam National University, Ho Chi Minh City 720300, Vietnam (e-mail: 1712213@student.hcmus.edu.vn).}
\thanks{Thuc D.~Nguyen is with the Faculty of Information Technology, University of Science, VNU-HCMC, Ho Chi Minh City 72711, Vietnam and Vietnam National University, Ho Chi Minh City 720300, Vietnam (e-mail: ndthuc@fit.hcmus.edu.vn).}}

\maketitle

\thispagestyle{plain}
\pagestyle{plain}


\IEEEpeerreviewmaketitle

\begin{abstract}
The goal of group testing is to efficiently identify a few specific items, called positives, in a large population of items via tests. A test is an action on a subset of items which returns positive if the subset contains at least one positive and negative otherwise. In non-adaptive group testing, all tests are fixed in advance and can be performed in parallel. In this work, we consider non-adaptive group testing with consecutive positives in which the items are linearly ordered and the positives are consecutive in that order. 

We present two contributions here. The first is the direct use of a binary code to construct measurement matrices compared to the use of Gray code in the state-of-the-art work, which is a rearrangement of the binary code, when the maximum number of consecutive positives is known. This leads to a reduction in decoding time in practice. The second one is efficient designs to identify positives when the number of consecutive positives is known. To the best of our knowledge, this setting has not been surveyed yet. Our simulations verify the efficiency of our proposed designs. In particular, it only requires up to $300$ tests to identify up to $100$ positives in a set of $2^{32} \approx 4.3\mathrm{B}$ items in less than $300$ nanoseconds. When the maximum number of consecutive positives is known, the simulations validate the superiority of our proposed design in decoding compared to the state-of-the-art work. Moreover, when the number of consecutive positives is known, the number of tests and the decoding time are almost reduced half.
\end{abstract}

\section{Introduction}
\label{sec:intro}

\subsection{Group testing}
\label{sec:intro:GT}

The goal of group testing (GT) is to efficiently identify up to $d$ positive items in a large population of $n$ items. Positive items satisfy some specific properties while negative items do not. Emerged by the seminal work of Dorfman~\cite{dorfman1943detection}, GT was considered as an efficient way to save time and money in identifying syphilitic draftees among a large population of draftees. Currently, with the ongoing Covid-19 pandemic since 2019, GT has been found to be an efficient tool for mass testing to identify infected persons~\cite{shental2020efficient,gabrys2020ac}. The strategy of GT is as follows. Instead of testing each item one by one to verify whether it is positive or negative, a group of items is pooled then tested. In the noiseless setting, the outcome of a test on a subset of items is positive if the subset has at least one positive and negative otherwise.

There are two basic approaches to designing tests. The first is \textit{adaptive} group testing in which the design of a test depends on the designs of the previous tests. This approach usually attains information-theoretic bounds on the number of tests required; however, it takes much time because of multiple stages. To remedy this drawback, the second approach, which is \textit{non-adaptive} group testing (NAGT), is to design all tests independently such that they can be performed simultaneously. NAGT can be represented by a measurement matrix in which an entry at row $i$ and column $j$ equals 1 indicates that the $j$th item belongs to test $i$ and that item does not belong to test $i$ otherwise. A test and an item are represented by a row and a column, respectively. NAGT has been widely applied in various fields such as computational and molecular biology~\cite{du2000combinatorial} and networking~\cite{DyachkovPSV19}. The focus of the work reported here is on the second approach, i.e., NAGT.

The procedure to get a measurement matrix is called \textit{construction} and the procedure to get the test outcomes by using the measurement matrix is called \textit{encoding}. The procedure to identify the positive items from the outcomes is called \textit{decoding}. Here we present two criteria on construction of a measurement matrix, which are the probability of obtaining a measurement matrix and the time and space it takes to generate a column in a measurement matrix. For the first criterion, a measurement matrix is random if part of the matrix or the whole matrix is obtained by chance, i.e., some tests are generated by a probabilistic scheme. In contrast, a measurement matrix is deterministic if every test is deterministic, i.e., every test is obtained with the probability of $1$. For the second criterion, a $t \times n$ measurement matrix is explicit if it takes time and space as a polynomial of the number rows and columns, i.e., $\poly(t, n)$, to generate a column in the matrix. A stricter condition is strongly explicit, in which it takes time and space as a polynomial of the number rows to generate a column in the matrix, i.e., $\poly(t, \log{n})$, where the number of rows is typically sublinear in $n$. From the two criteria, a measurement matrix is good for practice if it is deterministic and strongly explicit.

There are two main requirements to tackle group testing: minimize the number of tests and efficiently identify the set of positive items. The distribution of positives usually affects the encoding and decoding procedures of NAGT. There are two common distributions in group testing literature: probabilistic and combinatorial. In the \textit{probabilistic setting}, the distribution of positives is described by a probabilistic model. In the \textit{combinatorial setting}, any set of up to $d$ items can be the positive set. The number of tests required in the probabilistic setting is usually smaller than in the combinatorial setting. In the combinatorial setting, several schemes~\cite{du2000combinatorial,porat2011explicit,indyk2010efficiently,ngo2011efficiently,cheraghchi2013noise,bui2019efficient,cheraghchi2020combinatorial,cai2017efficient} have been proposed to attain a low number of tests, namely $O(d^{1 + o(1)} \log^{1 + o(1)}{n})$, and/or a low decoding time, namely, $\poly(d, \log{n})$. In the probabilistic setting, Cheraghchi and Nakos~\cite{cheraghchi2020combinatorial} and Price and Scarlett~\cite{price2020fast} presented schemes that achieve $O(d \log{n})$ tests to identify all positives in time $O(d \log{n})$. These results are an improvement of the previous result of Bondorf et al.~\cite{bondorf2021sublinear} in which the positives can be recovered by using $O(d \log{n})$ tests in time $O(d^2 \log{d} \cdot \log{n})$ with the decoding error probability vanishing as $n \rightarrow \infty$.

\subsection{Group testing with consecutive positives}
\label{sec:intro:GTwConsecutive}

In this work, we consider another distribution of positives, called \textit{the consecutive setting}, in which the input items are linearly ordered and the positives are consecutive in that order. Colbourn~\cite{colbourn1999group} first defined and called this specific setting of group testing \textit{group testing with consecutive positives}. This setting has been applied in genetic mapping and sequencing with linear DNA~\cite{balding1997design} or recently used in detecting infected individuals in connected communities~\cite{nikolopoulos2021group}.

Suppose that the positive items in the population of $n$ items are consecutive and the number of positives is up to $d$. Colbourn showed that the number of tests required can be reduced to $O(\log(dn))$ and $O \left( \log_2{ \frac{n}{d - 1} } + d \right)$ for adaptive and non-adaptive designs, respectively, which is much smaller than the bound $O(d^{1 + o(1)} \log^{1 + o(1)}{n})$ in the combinatorial setting. In addition, Juan and Chang~\cite{juan2008adaptive} could make the number of tests range from $\ceil{\log_2(dn)} - 1$ to $\ceil{ \log_2(dn)}$ in adaptive approach.

The focus of this work is on non-adaptive group testing with consecutive positives. For information-theoretic bound, Colbourn~\cite{colbourn1999group} showed that any group testing method must employ at least $\log_2(nd) - 1$ tests. On the other hand, he also showed the minimum number of tests required in any non-adaptive group testing with up to $d$ consecutive positives is $d - 1$. Therefore, the minimum number of tests in non-adaptive group testing with up to $d$ consecutive positives is $\max\{\log_2(nd) - 1, d - 1 \}$. Muller and Jimbo~\cite{muller2004consecutive} considered the case $d = 2$ and could construct an explicit measurement matrix with $\ceil{ \log_2{ \ceil{ \frac{n}{d - 1} } }} + 2d + 1$ tests. Unfortunately, they have not come up with an efficient decoding algorithm. Chang et al.~\cite{chang2015variation} later used random measurement matrices with $5 \log_2{\frac{n}{d}} + 2d + 21$ tests to identify all positives in time $O \left( \frac{n}{d} \log_2{\frac{n}{d}} + d^2 \right)$. In terms of both encoding and decoding, the seminal work of Colbourn~\cite{colbourn1999group} still remains the-state-of-the-art. In particular, he presented an efficient algorithm to identify all positives with $\ceil{\log_2{ \ceil{ \frac{n}{d - 1} } } } + 2d + 1$ tests in time $O \left( \log{\frac{n}{d - 1}} + d \right)$ where $d \geq 2$.

\subsection{Contributions}
\label{sub:intro:contri}

We consider two settings on the number of consecutive positives: i) the maximum number of consecutive positives is $d$ and ii) the number of consecutive positives is exactly $d$. We start with the first setting. Since neither Colbourn nor others analyze his decoding design in details, we first analyze it. A binary code is a collection of the binary representations of numbers expressed in the base-2 numeral system. Since Colbourn's encoding procedure uses Gray code, which is a rearrangement of the binary code, the outcome vector is composed by codewords in the Gray code. Note that a codeword of a code represents for an item and the decimal number corresponding to the codeword is the index of the item. Therefore, to identify the positives, one (directly or indirectly) needs to convert the codewords in the outcome vector from the Gray code to the binary code before getting the indices of the positives. From this observation, our first contribution is to directly use the binary code to remove the cost of converting from the Gray code to the binary code. This leads to a reduction in decoding time. In particular, although the number of tests in our design is slightly larger than in Colbourn's design, the decoding time in our design is smaller than in Colbourn's design in practice. The simulations in Section~\ref{sec:simul} verify our improvement.

To the best of our knowledge, the second setting in which the number of consecutive positives is known, has not been considered in existing literature. By using a design based on the Gray code, the number of required tests is $\ceil{\log_2{\ceil{\frac{n}{d}}}} + d$ which is almost $d + 1$ less than in Colbourn's design ($\ceil{\log_2{\ceil{\frac{n}{d - 1}}}} + 2d + 1$) when the maximum number of consecutive positives is $d$. On the other hand, if we use a design based on the binary code, the number of required tests is $2\ceil{\log_2{\ceil{\frac{n}{d}}}} + d$ which is still smaller than in Colbourn's design as long as $\log_2{\ceil{\frac{n}{d}}} < d + 1$.

The decoding complexity for both settings in our designs is $O \left( \log_2{\frac{n}{d}} + d \right)$. A summary of our comparison is shown in Table~\ref{tbl:comparison}.

\begin{table}[t]

\begin{center}
\scalebox{.785}{
\begin{tabular}{|l|c|c|c|c|c|c|}
\hline
\multicolumn{1}{|c|}{Scheme} & \begin{tabular}[c]{@{}c@{}}No. of\\ positives\end{tabular} & \begin{tabular}[c]{@{}c@{}}Design\\ approach\end{tabular} & \multicolumn{2}{c|}{\begin{tabular}[c]{@{}c@{}}Construction type of\\ measurement matrices\end{tabular}} & \begin{tabular}[c]{@{}c@{}}Number of tests\\ $t$\end{tabular} & \begin{tabular}[c]{@{}c@{}}Decoding time\\ (Decoding complexity)\end{tabular} \\ \hline
Colbourn~\cite{colbourn1999group} & \multirow{2}{*}{$\leq d$} & \multirow{2}{*}{Adaptive} & \multicolumn{2}{c|}{\multirow{2}{*}{Not available}} & $\ceil{\log_2(dn)} + c$ & \multirow{2}{*}{$t$ stages} \\ \cline{1-1} \cline{6-6}
Juan and Chang~\cite{juan2008adaptive} & & & \multicolumn{2}{c|}{}                                                                                    & $\ceil{\log_2(dn)} - 1 \leq t \leq \ceil{\log_2(dn)} + 1$ &                                                                               \\ \hline
Muller and Jimbo~\cite{muller2004consecutive} & $d = 2$ & \multirow{2}{*}{Non-adaptive} & \multicolumn{2}{c|}{Random, Explicit} & $\ceil{\log_2{ \ceil{\frac{n}{d - 1}}}} + 2d - 1$ & Not available                                                                 \\ \cline{1-2} \cline{4-7} 
Chang et al.~\cite{chang2015variation} & $\leq d$ & & \multicolumn{2}{c|}{Random, Explicit} & $5 \log_2{\ceil{\frac{n}{d}}} + 2d + 21$ & $O \left( \frac{n}{d} \log_2{\frac{n}{d}} + d^2 \right)$ \\ \hline
\multirow{2}{*}{Colbourn~\cite{colbourn1999group}} & $= 1$ & \multirow{5}{*}{Non-adaptive} & \multirow{5}{*}{\begin{tabular}[c]{@{}c@{}}Deterministic,\\ Strongly explicit\end{tabular}} & Binary code-based & $\ceil{\log_2{n}}$ & $O(\log{n})$ \\ \cline{2-2} \cline{5-7} 
& \begin{tabular}[c]{@{}c@{}}$\geq 2$ and $\leq d$\end{tabular} & & & Gray code-based & $\ceil{\log_2{ \ceil{ \frac{n}{d - 1} } } } + 2d + 1$ & $O \left( \log{\frac{n}{d - 1}} + d \right)$ \\ \cline{1-2} \cline{5-7} 
\textbf{Design in Theorem~\ref{thr:2decoding_UpToD}} & $\leq d$ & & & \textbf{Binary code-based} & $2 \ceil{\log_2{\ceil{\frac{n}{d}}}} + 2d$                    & \multirow{3}{*}{$O \left( \log_2{\frac{n}{d}} + d \right)$} \\ \cline{1-2} \cline{5-6}
\textbf{First design in Theorem~\ref{thr:2decoding_D}} & \multirow{2}{*}{$ = d$} & & & \textbf{Gray code-based} & $\ceil{\log_2{\ceil{\frac{n}{d}}}} + d + 3$ & \\ \cline{1-1} \cline{5-6} \textbf{Second design in Theorem~\ref{thr:2decoding_D}} & & & & \textbf{Binary code-based} & $2 \ceil{\log_2{\ceil{\frac{n}{d}}}} + d$ & \\ \hline
\end{tabular}}

\end{center}

\caption{Comparison of improved algorithms with previous ones. ``Not available'' means that the criterion does not hold or is not considered for that scheme. Parameter $c$ is some constant.}

\label{tbl:comparison}
\end{table}

\subsection{General idea of improved algorithms}
\label{sub:intro:sketch}

Although our improved algorithms are inspired by Colbourn's strategy~\cite{colbourn1999group}, we refine every technical details to attain efficient encoding and decoding procedures. Colbourn proposes a strategy to identify consecutive positives with two phases for $d \geq 2$. The first phase is to approximately locate where positives are. The outcome of this phase is a set of items, called \textit{potential positives}, that contains all positives and may contain some negatives. The second phase is to identify the true positives among the potential positives. In particular, in the first phase, the author partitions the $n$ (linearly ordered) items into subpools, in which we call them \textit{super items} here. A super item is a set of items which is positive if it contains at least one positive item and is negative otherwise. Each super item contains exactly $d - 1$ consecutive items and the last one may contain less than $d - 1$. Therefore, there are up to two consecutive super positive items that contain all positives. Colbourn uses Gray code as a measurement matrix to locate all super positive items. Based on the super positive item(s), he is able to locate exactly $d - 1$ or $2(d - 1)$ potential positives that contain all positives. In the second phase, there are $2(d - 1)$ tests and each test contains items spaced $2(d - 1)$ apart in the linear order. Hence, for any $2(d - 1)$ consecutive items, each test in the $2(d - 1)$ tests contains only one item of them. Therefore, by examining the outcome vector in the second phase, the true positives among the potential positives are precisely identified. The details of Colbourn's design can be found later in Section~\ref{sec:revisit}.

When the maximum number of consecutive positives is $d$, our improved algorithms are described here and more details with illustrations are presented in Section~\ref{sec:improved}. In the first phase, instead of distributing $d - 1$ consecutive items into each super item, we distribute $d$ consecutive items into each super item. Hence, there are up to two super positive items. Note that the last super item may contain less than $d$ items. Observe that adding $1$ to any number causes its binary representation to change as follows (assume the binary representation is not all $1$s): (i) find the left-most $0$ digit and flip it to $1$; (ii) flip all $1$s that appear after that digit to $0$. For instance, $01\mathbf{0}\underline{111}$ becomes $01\mathbf{1}\underline{000}$. We then use a binary code and its complement as a measurement matrix for the super items. This measurement matrix has the following properties: the numbers of ones in all columns are equal, and given an outcome vector, which is the union of two consecutive columns, the two columns can be identified based on the design of the measurement matrix and the observation. In the second phase, one creates $2d$ tests and each test contains items spaced $2d$ apart in the linear order. Because of this design, for any $2d$ consecutive items, a test contains only one item among them. Because there are up to two super positive items obtained from decoding the test outcomes in the first phase, there are up to $2d$ consecutive items that contain all positives. Therefore, by examining the test outcomes in the second phase, the true positives can be identified.

When the number of consecutive positives is exactly $d$, in the first phase, the $n$ items are partitioned to create $\ceil{n/d}$ super items in which each super item contains $d$ consecutive items. Note that the last super item may contain less than $d$ items. A measurement matrix used for testing these super items can be either generated from the Gray code as in Colbourn's design or the binary code and its complement as in the preceding paragraph. In the second phase, one creates only $d$ tests and each test contains items spaced $2d$ apart in the linear order. Depend on the construction choice of a measurement matrix in the first phase, we will have a corresponding decoding procedure. In particular, if the measurement matrix is based on the Gray code, the decoding procedure is as the same as Colbourn's one. On the other hand, if the measurement matrix is based on the binary code, the decoding procedure is as the same as the one in the first setting when the maximum number of consecutive positives is $d$. In the second phase, because there are exactly $d$ positives and each test contains items spaced $2d$ apart in the linear order, we can always identify either the smallest index of the positives (the starting positive) or the largest index of the positives (the terminal positive). Thus, all positives can be identified.

\section{Preliminaries}
\label{sec:pre}

Set of form $C = \{ c_1, \ldots, c_k \}$ used in this work is equipped with the linear order $c_i \prec c_{i + 1}$ for $1 \leq i < k$, where $\prec$ is the linear order notation. There are $n$ items indexed from 1 to $n$ to form set $N = [n] = \{1, 2, \ldots, n \}$.

\subsection{Super items}
\label{sub:pre:super}

\begin{center}
\begin{figure}[ht]
\centering
\includegraphics[scale=0.5]{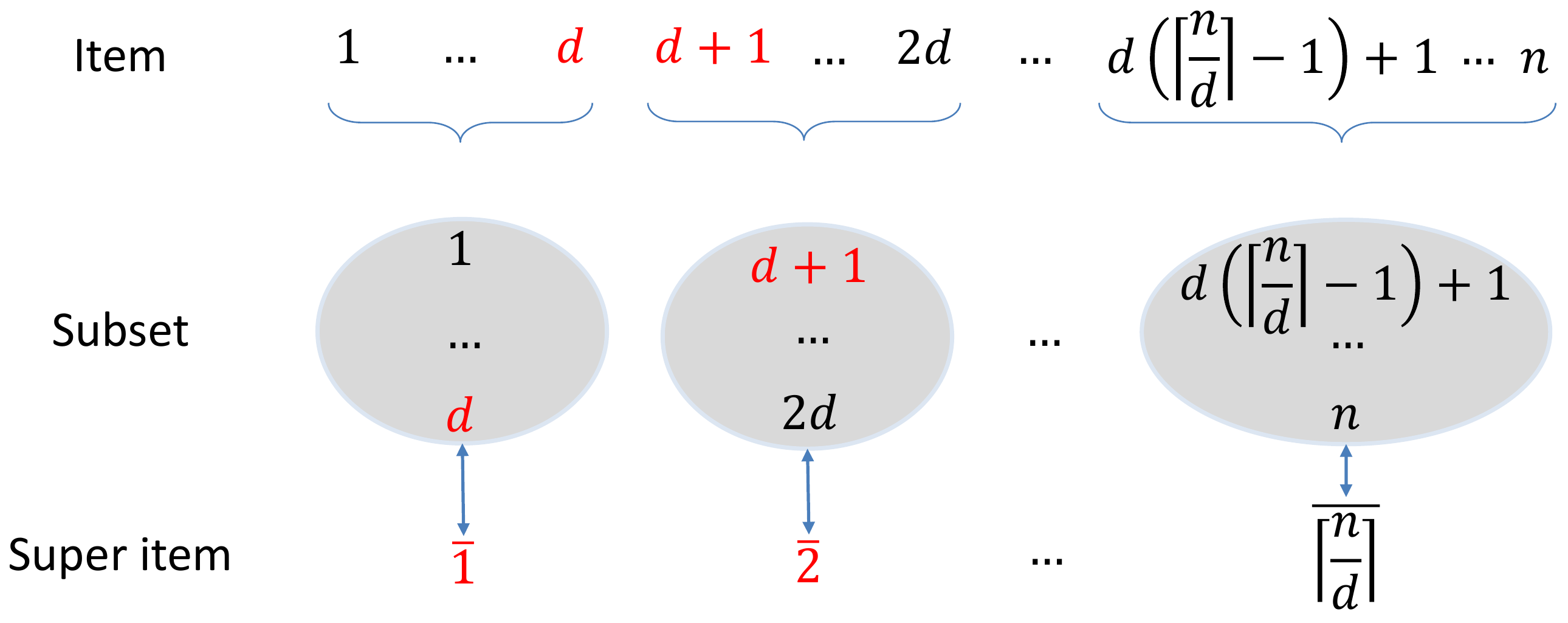}
\caption{Super items. A super item is a subset of items. Super items $\bar{1}$ and $\bar{2}$ are positive while the super item $\overline{\ceil{\frac{n}{d}}}$ is negative.}
\label{fig:SuperItem}
\end{figure}
\end{center}

We introduce the notion of super items here. A super item is a set of items which is positive if it contains at least one positive item and is negative otherwise. Suppose $\kappa = \ceil{n/d}$ super items are created from $n$ items in which the $j$th subset contains items indexed from $(j - 1)d + 1$ to $jd$ for $j = 1, \ldots, \ceil{n/d} - 1$, and the $\ceil{n/d}$th subset contains items indexed from $(\ceil{n/d} - 1)d + 1$ to $n$. The $j$th super item is the $j$th subset. An illustration of $\kappa$ super items is depicted in Fig.~\ref{fig:SuperItem}. Super items $\bar{1}$ and $\bar{2}$ are positive while the super item $\bar{\frac{n}{d}}$ is negative.

Given a $t \times \kappa$ measurement matrix $\cS = (s_{ij})$ used for $\kappa$ super items, the $j$th column represents for the $j$th super item. Entry $s_{ij}$ equals $1$ means all items belonging to super item $j$ are present in test $i$ and $s_{ij}$ equals $0$ means none of the items belonging in super item $j$ are present in test $i$. Then matrix $\cS$ can be enlarged to a $t \times n$ measurement matrix $\cS^\star$ used for the $n$ items, in which the $j$ column represents for the $j$th item, as follows: columns indexed from $(j - 1)d + 1$ to $jd$ in $\cM$ is assigned to be the $j$th column of $\cS$ ($\cS_j$). It is obvious that the outcome vector by using $\cS$ with the $\kappa$ super items is the same as the outcome vector by using $\cS^\star$ with the $n$ items.

\subsection{Notations}
\label{sub:pre:notation}

For consistency, we use capital calligraphic letters for matrices, non-capital letters for scalars, bold letters for vectors, and capital letters for sets. All matrix and vector entries are binary. The function $\toInt(\cdot)$ converts a vector from the binary representation to the decimal one and adds one. For example, if the input is $011$ then its index is $\toInt(011) = 0 \times 2^2 + 1 \times 2^1 + 1 \times 2^0 + 1 = 4$. For a set of consecutive positives $P = \{j_1, j_2, \ldots, j_{|P|} \}$, item $j_1$ and $j_{|P|}$ are called the starting and terminal positives, respectively. The main notations are as follows:

\begin{enumerate}
\item $n, d, \bX = (x_1, \ldots, x_n)^T$: number of items, (maximum) number of positive items, binary representation of $n$ items in which an entry $x_j = 1$ indicates that the $j$th item is positive and $x_j = 0$ indicates otherwise.
\item $P = \{j_1, j_2, \ldots, j_{|P|} \}$: set of positive items.
\item $\cT_{i, *}, \cT_{*, j}, \cS_j$: row $i$ of matrix $\cT$, column $j$ of matrix $\cT$, column $j$ of matrix $\cS$.
\item $\bV(i)$: the $i$th entry in the vector $\bV$.
\item $\log{x}$: the base-$2$ logarithm of $x$.
\end{enumerate}

\subsection{Problem definition}
\label{sub:pre:probDef}
Let $P$ be the positive set, where $|P| \leq d$. A test is defined by an action on a subset of $N$. The outcome of a test on a subset of items is positive if the subset contains at least one positive item, is negative otherwise.

We can model non-adaptive group testing with consecutive positives as follows. A $t \times n$ binary matrix $\cT =(t_{ij})$ is defined as a measurement matrix, where $n$ is the number of items and $t$ is the number of tests. Vector $\bX = (x_1,\ldots,x_n)^T$ is the binary representation vector of $n$ items, where $|\bX| = \sum_{j = 1}^n x_j \leq d$. If $|P| = a$ then there exists an index $0 \leq b \leq n - a$ such that $P = \{b + 1, b + 2, \ldots, b + a \}$. In other words, there exist two indices $a$ and $b$ such that $x_{b + 1} = x_{b + 2} = \ldots = x_{b + a} = 1$ and $x_j = 0$ for $j \in N \setminus \{b + 1, \ldots, b + a \}$, where $1 \leq a \leq d$ and $0 \leq b \leq n - a$. An entry $x_j=1$ indicates that item $j$ is positive, and $x_j=0$ indicates otherwise. The $j$th item corresponds to the $j$th column of the matrix. An entry $t_{ij}=1$ naturally means that item $j$ belongs to test $i$, and $t_{ij}=0$ means otherwise. The outcome of all tests is $\bY=(y_1, \ldots, y_t)^T$, where $y_i=1$ if test $i$ is positive and $y_i=0$ otherwise. Outcome vector $\bY$ is given by

\begin{equation}
\label{eqn:thresholdGT}
\bY = \cT \odot \bX = \begin{bmatrix}
\cT_{1, *} \odot \bX \\
\vdots \\
\cT_{t, *} \odot \bX
\end{bmatrix} = \begin{bmatrix}
y_1 \\
\vdots \\
y_t
\end{bmatrix}
\end{equation}
where $\odot$ represents the test operation in non-adaptive group testing; namely, $y_i = \cT_{i, *} \odot \bX = 1$ if $\sum_{j=1}^n x_j t_{ij} \geq 1$ and $y_i = \cT_{i, *} \odot \bX = 0$ if $\sum_{j=1}^n x_j t_{ij} = 0$ for $i = 1, \ldots, t$.

Our objective is to find an efficient encoding and decoding design to identify up to $d$ consecutive positives in non-adaptive group testing with consecutive positives. Precisely, our task is to minimize the number of rows in matrix $\cT$ and the time for recovering $\bX$ from $\bY$ by using $\cT$.

\section{Colbourn revisited}
\label{sec:revisit}

\subsection{Overview}
\label{sub:revisit:overview}

\begin{figure}[ht]
\centering
\includegraphics[scale=0.435]{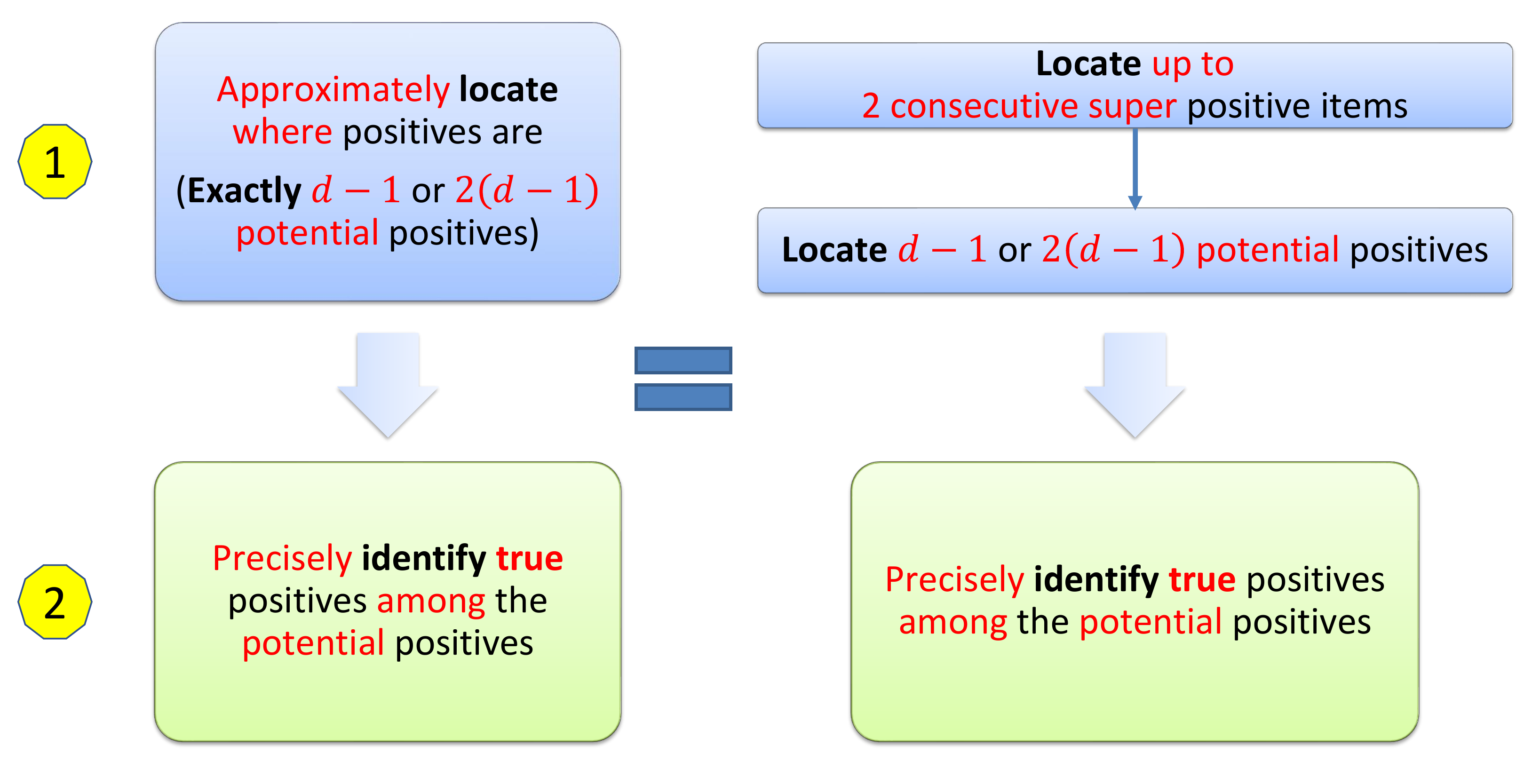}
\caption{Colbourn's design.}
\label{fig:Colbourn}
\end{figure}

Since our proposed designs are based on Colbourn's strategy~\cite{colbourn1999group}, we start revisiting it here. However, the decoding procedure of Colbourn's design has not been analyzed by the author or others, our minor contribution is to make its full analysis. Colbourn proposes a strategy to identify consecutive positives in two phases. The strategy is illustrated in Fig.~\ref{fig:Colbourn}. The first phase is to approximately locate where positives are. The outcome of this phase is a set of items, called \textit{potential positives}, that contains all positives and some false positives. The second phase is to identify the true positives among the potential positives. In particular, the author partitions the $n$ (linearly ordered) items into \textit{super items} as defined in Section~\ref{sub:pre:super} and each super item contains exactly $d - 1$ items. The last super item may contain less than $d - 1$ consecutive items. Therefore, there are up to two consecutive super positive items that contain all positives. Based on the super positive item(s), it is possible to identify exactly $d - 1$ or $2(d - 1)$ potential positives. This phase is illustrated in the rectangles with rounded corners and light blue background in (the left side of) Fig.~\ref{fig:Colbourn}. The second phase is to identify all positives among the potential positives and is illustrated in the rectangles with rounded corners and light green background in (the right side of) Fig.~\ref{fig:Colbourn}.

\begin{figure}[ht]
\centering
\includegraphics[scale=0.45]{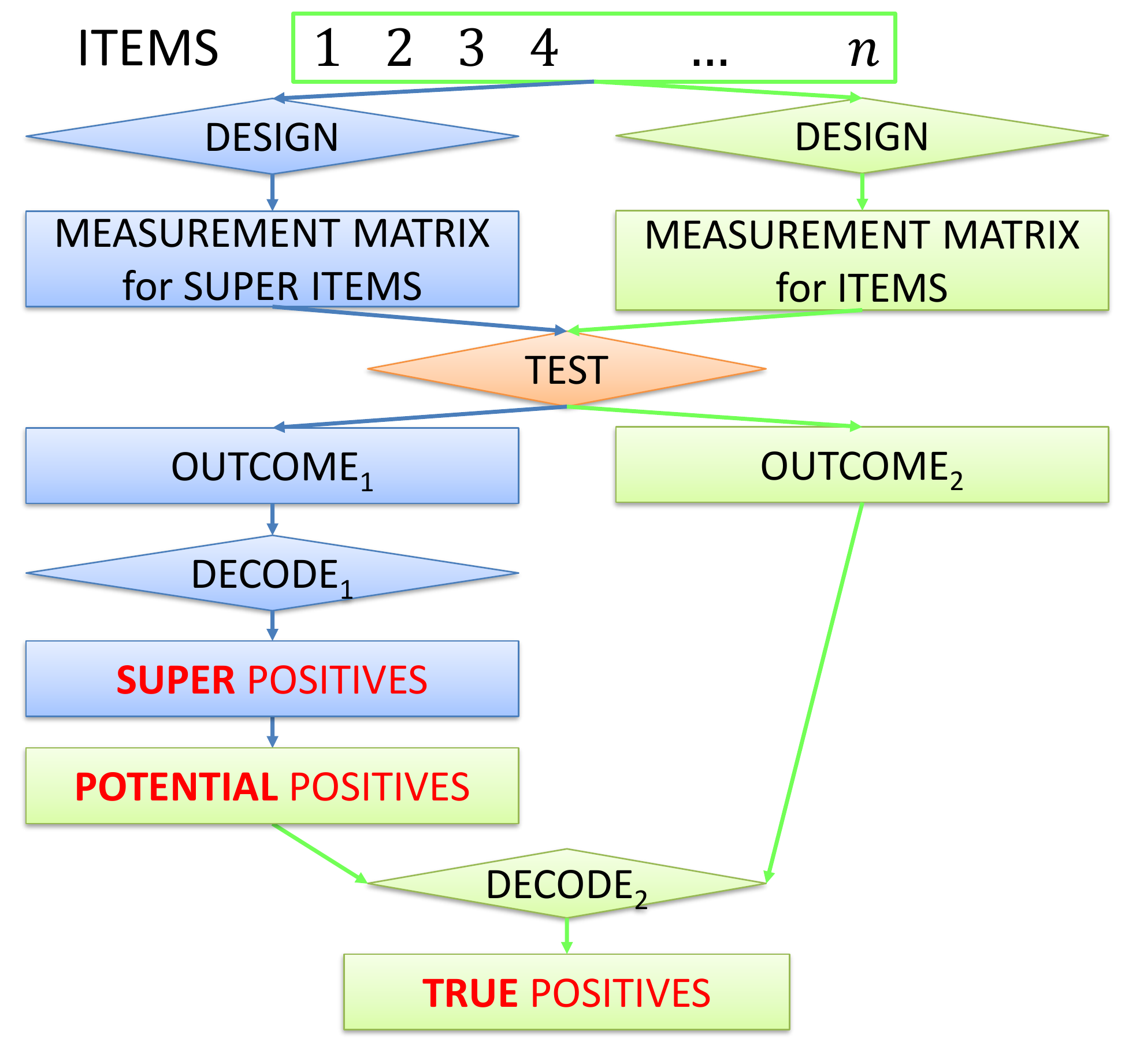}
\caption{Details of Colbourn's design.}
\label{fig:Colbourn_details}
\end{figure}

The details of Colbourn's design are illustrated in Fig.~\ref{fig:Colbourn_details}. Flow is from top to bottom. The flow of each phase is followed by consistent arrow color. In particular, the first and second phases include drawings followed by blue and light green arrows, respectively. Although the measurement matrix in the first phase, denoted as $\cT_1$, is designed for super items and the measurement matrix in the second phase, denoted as $\cT_2$, is designed for items, testing can be performed simultaneously for both phases. However, one needs potential positives as the output of the first phase to combine with the outcome vector in the second phase in order to finally identify the true positives, i.e., the items in $P$. The decoding complexity of Colbourn's design is equal to the complexity of finding potential positives in the first phase plus the complexity of identifying the true positives among the potential positives obtained in the first phase.

The results on the non-adaptive strategy of Colbourn's design can be summarized as follows:
\begin{theorem}~\cite[Extended version of Theorem 3.2]{colbourn1999group}
In a linearly ordered set of $n$ items with at least one positive and at most $d$ consecutive positives, there exists a deterministic and strongly explicit measurement matrix with $t = \ceil{\log_2{ \ceil{ \frac{n}{d - 1} } } } + 2d + 1$ tests such that all positives can be found in $O(t) = O(\log_2{\frac{n}{d - 1}} + d)$ time.
\label{thr:ColbournX}
\end{theorem}

\subsection{Encoding procedure}
\label{sub:revisit:enc}

Colbourn treats the case when the number of positives is exactly $1$, i.e., $d = |P| = 1$, separately. He forms $\ceil{\log_2{n}}$ tests in which the binary representation vector of $j - 1$ represent for item $j$. The $i$th test consists of all items $j$ for which the $i$th bit in the binary representation of $j$ is $1$.

From now, he assumes $|P| \geq 2$. This implies $d \geq 2$. Colbourn designs three separate measurement matrices in which two of them are used in the first phase and the remaining matrix is used in the second phase. The $n$ items are partitioned into $\kappa = \ceil{n/(d - 1)}$ super items in which the $j$th super item contains $(d - 1)$ consecutive items index from $(j - 1)(d - 1) + 1$ to $j(d - 1)$. Note that the last super item may contain less than $d - 1$ items. The measurement matrix $\cT_1$ in the first phase consists of two matrices $\cS$ and $\cL$, i.e., $\cT_1 = \begin{bmatrix}
\cS \\ \cL
\end{bmatrix}$, while the measurement matrix $\cT_2$ in the second phase consists of only one measurement matrix $\cV$, i.e., $\cT_2 = \cV$. The details are as follows.

Set $s = \ceil{\log_2{\kappa}}$. Let $\cS$ be an $s \times \kappa$ measurement matrix created from an $s$-bit Gray code~\cite{kreher2020combinatorial}. An $s$-bit Gray code is an ordering of the $2^s$ binary vectors of length $s$ such that any two consecutive vectors differ in only one position. For example, when $n = 16$ and $d = 5$ ($\kappa = 4$), the following is a $2 \times 4$ matrix $\cS$:
\begin{equation}
\cS = \left[ \begin{array}{cccc}
0 & 0 & 1 & 1 \\
0 & 1 & 1 & 0
\end{array} \right]. \nonumber
\end{equation}

Since there exists a deterministic decimal-to-Gray conversion~\cite{Irshid87} that takes $O(s)$ time to convert a decimal value to an $s$-bit vector in the Gray code, any column in $\cS$ is deterministic and can be generated in $O(s)$ time.

The second matrix of size $3 \times \kappa$, denoted as $\cL = (l_{ij})$, is created by assigning entry $l_{ij} = 1$ if $j \equiv i \Mod{3}$ and $l_{ij} = 0$ otherwise. In other words, each test contains items spaced $3$ apart in the linear order. For example, when $n = 16$ and $d = 5$, the following is a $3 \times 4$ matrix $\cL$:
\begin{equation}
\cL = \left[ \begin{array}{cccc}
1 & 0 & 0 & 1 \\
0 & 1 & 0 & 0 \\
0 & 0 & 1 & 0
\end{array} \right]. \nonumber
\end{equation}

The last matrix of size $2(d - 1) \times n$, denoted as $\cV = (v_{ij})$, is created by assigning entry $v_{ij} = 1$ if $j \equiv i \Mod{(2(d - 1))}$ and $v_{ij} = 0$ otherwise. In other words, each test contains items spaced $2(d - 1)$ apart in the linear order. For example, when $n = 16$ and $d = 5$, the following is a $8 \times 16$ matrix $\cV$:
\begin{equation}
\cV = \left[ \begin{array}{cccccccc|cccccccc}
1 & 0 & 0 & 0 & 0 & 0 & 0 & 0 & 1 & 0 & 0 & 0 & 0 & 0 & 0 & 0 \\
0 & 1 & 0 & 0 & 0 & 0 & 0 & 0 & 0 & 1 & 0 & 0 & 0 & 0 & 0 & 0 \\
0 & 0 & 1 & 0 & 0 & 0 & 0 & 0 & 0 & 0 & 1 & 0 & 0 & 0 & 0 & 0 \\
0 & 0 & 0 & 1 & 0 & 0 & 0 & 0 & 0 & 0 & 0 & 1 & 0 & 0 & 0 & 0 \\
0 & 0 & 0 & 0 & 1 & 0 & 0 & 0 & 0 & 0 & 0 & 0 & 1 & 0 & 0 & 0 \\
0 & 0 & 0 & 0 & 0 & 1 & 0 & 0 & 0 & 0 & 0 & 0 & 0 & 1 & 0 & 0 \\
0 & 0 & 0 & 0 & 0 & 0 & 1 & 0 & 0 & 0 & 0 & 0 & 0 & 0 & 1 & 0 \\
0 & 0 & 0 & 0 & 0 & 0 & 0 & 1 & 0 & 0 & 0 & 0 & 0 & 0 & 0 & 1
\end{array} \right]. \nonumber
\end{equation}

In summary, the numbers of tests in the first and second phases are $s + 3$ and $2(d - 1)$, respectively, where $s = \ceil{\log_2{\kappa}}$ and $\kappa = \ceil{n/(d - 1)}$. Hence, the total number of tests in the two phases is $s + 3 + 2(d - 1) = \ceil{\log_2{\ceil{\frac{n}{d - 1}}}} + 2d + 1$. Moreover, since every column in $\cS$, $\cL$, $\cV$ is deterministic and can be generated in time $O(s)$, $O(1)$, and $O(d)$, the measurement matrices used in the encoding procedure are deterministic and strongly explicit.

\subsection{Decoding procedure}
\label{sub:revisit:dec}

Let $\bY_1$ and $\bY_2$ be the outcome vectors obtained by using $\cT_1$ and $\cT_2$, respectively. The decoding procedure to obtain the set of super positive items from decoding $\bY_1$ is denoted as $\mathrm{Dec2ConsecutivePositives}_{\mathrm{Gray}}(\bY_1, \cT_1)$.

Let $\bS$, $\bL$, and $\bV$ be the outcome vectors by using the measurement matrices $\cS$, $\cL$, and $\cV$. Vector $\bS$ and matrix $\cL$ are used to identify one super positive item. It has known that there are up to two super positive items because of the generating of the $\kappa$ super items. We have the column in the measurement matrix corresponding to a super positive item is always included in the outcome vector. Because $\cS$ is a Gray code, any two consecutive columns differ in only one position. Therefore, the union of two consecutive columns is always one of them. This implies the outcome vector $\bS$ is always identical to the column in $\cS$ corresponding to one super positive item regardless of the number of super positive items. By converting $\bS$ from the Gray code to a decimal number, one gets the index $\alpha$ of a super positive item.

Even the index of a super positive item is identified, it remains unknown one or two super positive items are present. Moreover, when there are exactly two super positive items, it is undecided to claim the preceding super item of $\bar{\alpha}$, i.e., $\overline{\alpha - 1}$, or the succeeding super item of $\bar{\alpha}$, i.e., $\overline{\alpha + 1}$, as the super positive item. To resolve these problems, vector $\bL$ and matrix $\cL$ are used to to identify how many super positive items are present and the index of the remaining super positive item (if available) in this case. Because of the design of $\cL$, each of three consecutive super items appears in exactly one of the three tests in $\cL$, and when it appears in a test, the other two super items do not. Hence, by checking $\bL$, one can determine exactly all super positive items. 

After executing the decoding procedure for the first two outcome vectors $\bS$ and $\bL$, one or two super positive items that contain all positives are identified. Therefore, we obtain a set of $d - 1$ or $2(d - 1)$ potential positives. Because of the structure of $\cV$, for any $2(d - 1)$ consecutive items, each test in the last $2(d - 1)$ tests contains only one item of them. Thus, by examining vector $\bV$, the true positives among the potential positives are precisely identified.

\subsection{Decoding complexity}
\label{sub:revisit:complexity}

Since it takes $O(s)$ time to convert an $s$-bit vector in Gray code to an $s$-bit vector in another binary representation~\cite{Irshid87} and another $O(s)$ time to convert an $s$-bit vector in the binary representation to its corresponding decimal value, the time complexity of converting an $s$-bit vector in Gray code to its corresponding decimal number is $O(s) + O(s) = O(s)$.

In the first phase, one needs to convert the outcome vector $\bS$ in the $s$-bit Gray code to a decimal value $\alpha$. This step takes $O(s)$ time as analyzed in the preceding paragraph. Then examining $\bL$ to locate another super positive item if available takes $O(3) = O(1)$ time. Therefore, the complexity of the first phase is $O(s) + O(1) = O(s)$. In the second phase, for any item, we can explicitly locate which test it belongs among the last $2(d - 1)$ tests. In particular, item $j$ belongs to test $i$ if and only if $j \equiv i \Mod{2(d - 1)}$ for $i = 1, 2, \ldots, 2(d - 1)$. Hence, by scanning the outcome vector $\bV$ one time with the knowledge of $(d - 1)$ or $2(d - 1)$ potential positives, the true positives can be identified. In other words, the complexity of the second phase is $O(2(d - 1)) = O(d)$. In summary, the decoding complexity of Colbourn's design is $O(s) + O(d) = O(s + d) = O(\log_2{(n/(d-1))} + d)$.

\section{Identifying of up to two consecutive positives}
\label{sec:2Positives}

\subsection{Overview}
\label{sub:2Positives:overview}

To efficiently identify up to two super positive items in the first phase of Colbourn's design, we present a non-adaptive design for identifying up to two consecutive positives among $n$ items. The basic idea of our proposed design is to exploit the structure of a deterministic and strongly explicit measurement matrix and the linear order of $n$ items. We create a deterministic and strongly explicit measurement matrix such that the union of any two consecutive columns in it is different from the union of other two consecutive columns. Based on this property and the measurement matrix structure, we carefully develop a decoding scheme whose decoding time is linear to the number of measurements. The encoding and decoding procedures to identify up to two consecutive positives are described in Algorithm~\ref{alg:2Positives:Dec1} and summarized in Lemma~\ref{lem:2decoding}.

\begin{lemma}
Let $n$ be a positive integer and $N = \{ 1, 2, \ldots, n \}$ be the set of $n$ linearly ordered items. Then there exists a deterministic and strongly explicit $2 \ceil{\log_2{n}} \times n$ measurement matrix such that if $N$ has up to two consecutive positives, the positives can be identified with $s = 2 \ceil{\log_2{n}}$ tests in $O(s) = O(\log_2{n})$ time.
\label{lem:2decoding}
\end{lemma}

\subsection{Encoding procedure}
\label{sub:2Positives:enc1}

Let $\cS$ be an $s \times n$ measurement matrix associated with the input set of items $N = \{1, 2, \ldots, n \}$:
\begin{equation}
\cS = \begin{bmatrix}
\bB_1 & \bB_2 & \ldots & \bB_n \\
\overline{\bB}_1 & \overline{\bB}_2 & \ldots & \overline{\bB}_n
\end{bmatrix} =
\begin{bmatrix}
\cS_1 & \ldots & \cS_n
\end{bmatrix}, \label{matrixS}
\end{equation}
where $s = 2 \left\lceil \log{n} \right\rceil$, $\bB_j$ is the $\left\lceil \log{n} \right\rceil$-bit binary representation of integer $j-1$, $\overline{\bB}_j$ is the complement of $\bB_j$, and $\cS_j := \begin{bmatrix} \bB_j \\ \overline{\bB}_j \end{bmatrix}$ for $j = 1,2,\ldots, n$. Column $\cS_j$ represents for the $j$th item of $N$ and that the weight of every column in $\cS$ is $s/2 = \left\lceil \log{n} \right\rceil$. Furthermore, the $j$th item of $N$, which is also item $j$, is uniquely identified by $\bB_j$. For example, if we set $n = 8$, $s = 2 \log{n} = 6$, and the matrix in~\eqref{matrixS} becomes:
\begin{align}
\cS = \begin{bmatrix}
0 & 0 & 0 & 0 & 1 & 1 & 1 & 1 \\
0 & 0 & 1 & 1 & 0 & 0 & 1 & 1 \\
0 & 1 & 0 & 1 & 0 & 1 & 0 & 1 \\
1 & 1 & 1 & 1 & 0 & 0 & 0 & 0 \\
1 & 1 & 0 & 0 & 1 & 1 & 0 & 0 \\
1 & 0 & 1 & 0 & 1 & 0 & 1 & 0 \\
\end{bmatrix}. \label{exampleS}
\end{align}

Let $\bS = (s_1, \ldots, s_n)^T$ be the binary representation vector of $n$ items with up to two consecutive positives, where an entry $s_j = 1$ indicates that item $j$ is positive, and $s_j = 0$ indicates otherwise. Vector $\bS$ satisfies the following properties:
\begin{itemize}
\item $|\bS| \leq 2$.
\item When $|\bS| = 2$, there exists an index $b$ such that $s_{b + 1} = s_{b + 2} = 1$ and $s_j = 0$ where $j \in N \setminus \{b + 1, b + 2 \}$ for $0 \leq b \leq n - 2$.
\end{itemize}

The outcome vector by performing tests on the input set of items $N$ and its measurement matrix $\cS$ is $\bY = \cS \odot \bS$.

\subsection{Decoding procedure}
\label{sub:2Positives:dec1}

The decoding procedure is summarized in Algorithm~\ref{alg:2Positives:Dec1}. Step~\ref{alg:2Positives:left_most} is first to divide the outcome vector $\bY^T$ into equal vectors, which are $\bY_L$ (the left half of $\bY^T$) and $\bY_R$ (the right half of $\bY^T$), and find the left-most $1$ digit of $\bY_L$ such that $\bY_L(i_0) = \bY_R(i_0) = 1$. If such left-most $1$ digit exists, there are two positives in the input set of items. Based on the property of the union of two consecutive columns in $\cS$, this step is then to recover the indices of the two positives by setting $\bY_L(i_0) = 0$ and returning $\{ \toInt(\bY_L), \toInt(\bY_L) + 1 \}$. If such $i_0$ does not exist, Step~\ref{alg:2Positives:1Pos} proceeds to return the index of the only positive or an empty set.

\begin{algorithm}
\caption{$\mathrm{Dec2ConsecutivePositives}_{\mathrm{Bin}}(\bY, \cS)$: Decoding procedure for group testing with up to two consecutive positives.}
\label{alg:2Positives:Dec1}
\textbf{Input:} Outcome vector $\bY$, matrix $\cS$ of size $s \times n$ defined in~\eqref{matrixS}.\\
\textbf{Output:} Set of up to two consecutive positives.

\begin{algorithmic}[1]
\State Write $\bY^T = (\bY_L, \bY_R)$, and find the position $i_0$ of the left-most $1$ digit of $\bY_L$ such that $\bY_L(i_0) = \bY_R(i_0) = 1$. If $i_0$ exists, set $\bY_L(i_0) = 0$ and return $\{ \toInt(\bY_L), \toInt(\bY_L) + 1 \}$ \label{alg:2Positives:left_most}
\State Otherwise, return $\{ \toInt(\bY_L) \}$ if there exists $i$ such that $\bY_L(i) \vee \bY_R(i) = 1$ and return an empty set if such $i$ does not exist. \label{alg:2Positives:1Pos}
\end{algorithmic}
\end{algorithm}

\subsection{Correctness and the decoding complexity}
\label{sub:2Positives:correctness1}

To prove that matrix $\cS$ in~\eqref{matrixS} can be used to identify two positives which are consecutive or up to one positive in a population of $n$ linearly ordered items, we first state the following lemma.

\begin{lemma}
Given matrix $\cS$ defined in~\eqref{matrixS}, for any two distinct indices $a \neq b$ in $[n - 1]$, we have $\cS_a \vee \cS_{a + 1} \not\equiv \cS_b \vee \cS_{b + 1}$ and $\cS_a \vee \cS_{a + 1} \not\equiv \cS_b$.
\label{lem:2Unique}
\end{lemma}

\begin{proof}

Observe that adding $1$ to any number causes its binary representation to change as follows (assume the binary representation is not all $1$s): (i) find the left-most $0$ digit and flip it to $1$; (ii) flip all $1$s that appear after that digit to 0. For instance, $01\mathbf{0}\underline{111}$ becomes $01\mathbf{1}\underline{000}$.

Now, when we take the bit-wise OR of the representations of $a$ and $a + 1$, the outputs at flipped locations will clearly all be of the form $0 \vee 1$ or $1 \vee 1$, i.e., always $1$, whereas the non-flipped locations will be $0 \vee 0$ or $1 \vee 1$. Hence, since each column of $\cS$ contains both the binary string and its complement, we can conclude for $\bA = \cS_a \vee \cS_{a + 1}$ and $i \in \{1, 2, \ldots, s/2 \}$ that $\bA(i) = \bA(i + s/2)$ ($= 1$) if and only if the $i$-th bit corresponds to one of the flipped bits. Hence, $\cS_a \vee \cS_{a + 1}$ is equal to $\cS_b \vee \cS_{b + 1}$ if and only if $a$ equals to $b$.

Regarding the case $\cS_a \vee \cS_{a + 1} \not\equiv \cS_b$, we can prove this by comparing the weights of $\cS_a \vee \cS_{a + 1}$ and $\cS_b$. Because $\cS_a \neq \cS_{a + 1}$, we must have $\wt(\cS_a \vee \cS_{a + 1}) \geq s/2 + 1$, where $\wt(\cdot)$ is the number of ones in the input vector. On the other hand, the weight of $\cS_b$ is always $s/2$. Therefore, we imply that $\cS_a \vee \cS_{a + 1} \not\equiv \cS_b$.
\end{proof}

We are now ready to prove the correctness of Algorithm~\ref{alg:2Positives:Dec1}. If there exists the position $i_0$ of the left-most $1$ digit of $\bY_L$ such that $\bY_L(i_0) = \bY_R(i_0) = 1$, there must be exactly two positives. Otherwise, we get $\bY_L = 1 - \bY_R$ when there is only one positive and we get $\bY_L = \bY_R = 0$ when there are no positives. This implies the condition $\bY_L(i_0) = \bY_R(i_0) = 1$ never happens when there is less than two positives. Because of the proof of Lemma~\ref{lem:2Unique} for the case $\cS_a \vee \cS_{a + 1} \neq\cS_b \vee \cS_{b + 1}$ when $a \neq b$, if there exists the left-most $1$ digit of $\bY_L$ such that $\bY_L(i_0) = \bY_R(i_0) = 1$, there are two positives in the input set, says $a$ and $a + 1$. Let us denote the position of the left-most $1$ digit of $\bY_L$ as $i_0$. This position is also the position of the left-most $0$ digit of the binary representation of $a$ ($\bB_a$). Therefore, Step~\ref{alg:2Positives:left_most} is to retrieve that representation and return the two positives. Once there does not exist the left-most $1$ digit of $\bY_L$ such that $\bY_L(i_0) = \bY_R(i_0) = 1$, there is one positive or there are no positives in the input set $N$. Step~\ref{alg:2Positives:1Pos} is hence to return $\{ \toInt(\bY_L) \}$ when the input set contains only one positive and to return an empty set when there are no positives in the input set.

Matrix $\cS$ is obviously deterministic and strongly explicit because the $j$th column of $\cS$ is the $\lceil \log{n} \rceil$-big binary representation of integer $j - 1$. To complete both Steps~\ref{alg:2Positives:left_most} and~\ref{alg:2Positives:1Pos}, it only requires to scan $\bY_L$ once, i.e., it takes $O(s/2)$ time to complete the scanning. Moreover, since converting an $s/2$-bit binary vector to its corresponding decimal number takes $O(s/2)$ time, the decoding complexity of Algorithm~\ref{alg:2Positives:Dec1} is thus $O(2s/2) = O(s)$.

\subsection{Example for Algorithm~\ref{alg:2Positives:Dec1}}
\label{sub:2Positives:eg}

Set $n = 8$. Then a measurement matrix $\cS$ of size $6 \times 8$ can be constructed as in~\eqref{exampleS}. Consider four sets of positives $P_1 = \emptyset, P_2 = \{ 1 \}, P_3 = \{1, 2 \}$, and $P_4 = \{4, 5 \}$. Their corresponding outcome vectors by using $\cS$ as a measurement matrix are:
\begin{equation}
\bY_1 = \begin{bmatrix}
0 \\
0 \\
0 \\
0 \\
0 \\
0
\end{bmatrix}, \bY_2 = \begin{bmatrix}
0 \\
0 \\
0 \\
1 \\
1 \\
1
\end{bmatrix}, \bY_3 = \begin{bmatrix}
0 \\
0 \\
1 \\
1 \\
1 \\
1
\end{bmatrix}, \bY_4 = \begin{bmatrix}
1 \\
1 \\
1 \\
1 \\
1 \\
1
\end{bmatrix}.
\end{equation}

We now start decoding $\bY_1, \bY_2, \bY_3$, and $\bY_4$. For $\bY_1^T = (\bY_L^1, \bY_R^1)$, there does not exist $i_0$ such that $\bY_L^1(i_0) = \bY_R^1(i_0) = 1$. Therefore, Algorithm~\ref{alg:2Positives:Dec1} proceeds to Step~\ref{alg:2Positives:1Pos}. Because there does not exist $i$ such that $\bY_L^1(i) \vee \bY_R^1(i) = 1$, Step~\ref{alg:2Positives:1Pos} returns an empty set which coincides with $P_1$. For $\bY_2^T = (\bY_L^2, \bY_R^2)$, since there does not exist $i_0$ such that $\bY_L^2(i_0) = \bY_R^2(i_0) = 1$, Algorithm~\ref{alg:2Positives:Dec1} proceeds to Step~\ref{alg:2Positives:1Pos}. Because $\bY_L^2(1) \vee \bY_R^2(1) = 1$, this step returns $\{ \toInt(\bY^1_L) \} = \{ 1 \} = P_2$.

We now examine $\bY_3$ and $\bY_4$. For $\bY_3^T = (\bY_L^3, \bY_R^3)$, since $\bY_L^3(3) = \bY_R^1(3) = 1$, Step~\ref{alg:2Positives:left_most} changes $\bY_L^3$ to be $(0, 0, 0)$ by setting $\bY_L^3(3)$ to be zero. Then, the set of two positives is $\{ \toInt(\bY^3_L), \toInt(\bY^3_L) + 1 \} = \{1, 2 \} = P_3$. Similarly, for $\bY_4^T = (\bY_L^4, \bY_R^4)$, since $\bY_L^4(1) = \bY_R^4(1) = 1$, Step~\ref{alg:2Positives:left_most} changes $\bY_L^4$ to be $(0, 1, 1)$ by setting $\bY_L^4(1)$ to be zero. It then returns the set of two positives $\{ \toInt(\bY^4_L), \toInt(\bY^4_L) + 1 \} = \{4, 5 \} = P_3$.

\section{Improved algorithm for group testing with the maximum number of consecutive positives}
\label{sec:improved}

Although our proposed design is based on Colbourn's design, we replace the measurement matrices used in that design with ours. The result of our improved algorithm is summarized as follows.

\begin{theorem}
Let $n$ be a positive integer and $N = \{ 1, 2, \ldots, n \}$ be the set of linearly ordered items with up to $d$ consecutive positives. Then there exists a deterministic and strongly explicit measurement matrix such that the positives can be identified with $2 \lceil \log_2{\frac{n}{d}} \rceil + 2d$ tests in $O \left( \log_2{\frac{n}{d}} + d \right)$ time.
\label{thr:2decoding_UpToD}
\end{theorem}

\subsection{Encoding procedure}
\label{sub:algs:enc}

The encoding procedure includes the first and seconds phases as described in Section~\ref{sub:revisit:enc}. We first create $\kappa = \ceil{n/d}$ \textit{super items} with linear order in which each super item contains exactly $d$ items, except for the last super item which may contain less than $d$ items. In particular, the $n$ items are distributed into $\ceil{n/d}$ subsets and the $j$th subset contains items indexed from $(j - 1)d + 1$ to $jd$. Note that the last super item may contain less than $d$ items. The $j$th super item is the $j$th subset. When the last super item is positive, it is obvious there are up to two consecutive super positive items. Otherwise, we also have there are up to two consecutive super positive items because the input items are linearly ordered, the number of positive items is up to $d$, the positive items are consecutive and each super item contains exactly $d$ items. 

In the first phase, the measurement matrix $\cT_1 = \cS = (s_{ij})$ used here is as the same as the one in~\eqref{matrixS} by replacing items with super items and $n$ with $\ceil{n/d}$. Every item in super item $j$ belongs to test $i$ if and only if $s_{ij} = 1$. The outcome vector by using $\cT_1$ is $\bY_1$. Let $\bar{\bX} = (x_{\bar{1}}, \ldots, x_{\bar{\kappa}})^T$, where $x_{\bar{j}} = 1$ if and only if the super item $\bar{j}$ is positive, and $x_{\bar{j}} = 0$ otherwise. Since any two super items do not share an item, items indexed from $(j - 1)d + 1$ to $jd$ belong to test $i$ if and only if $s_{ij} = 1$. Therefore, we obtain $\bY_1 = \cS \odot \bar{\bX}$. Note that we also have $\bY_1 = \cS^\star \odot \bX$, where $\cS^\star$ is defined in Section~\ref{sub:pre:super}.

In the second phase, a $2d \times n$ measurement matrix $\cT_2 = \cH = (h_{ij})$ is created as follows: entry $h_{ij}$ equals to $1$ if and only if $j \equiv i \Mod(2d)$. In other words, each test contains items spaced $2d$ apart in the linear order. The outcome vector by using $\cT_2$ is $\bY_2 = \cH \odot \bX$.

The final outcome vector by using both $\cT_1$ and $\cT_2$ for the first and second phases is $\bY = \begin{bmatrix} \bY_1 \\ \bY_2 \end{bmatrix} = \begin{bmatrix} \cS^\star \\ \cT_2 \end{bmatrix} \odot \bX$. Since the numbers of rows in $\cT_1$ and $\cT_2$ are $2 \ceil{\log_2{\ceil{\frac{n}{d}}}}$ and $2d$, respectively, the number of tests for this design is $2 \ceil{\log_2{\ceil{\frac{n}{d}}}} + 2d$.

\subsection{Decoding procedure and the decoding complexity}
\label{sub:algs:dec}

Similar to Colbourn's decoding procedure, in the first phase, we start decoding $\bY_1$ to identify super positive items and then infer (up to $2d$) potential positives. Because of the design of $\cH$, for any $2d$ consecutive items, a test induced from $\cH$ contains only one item among them. Moreover, since there are up to $2d$ potential positives, by examining $\bY_2$, the true positives can be identified. This decoding procedure is described in Algorithm~\ref{alg:ImprovedUpToD}.

\begin{algorithm}
\caption{Decoding procedure for group testing with up to $d$ consecutive positives.}
\label{alg:ImprovedUpToD}
\textbf{Input:} Outcome vector $\bY = \begin{bmatrix} \bY_1 \\ \bY_2 \end{bmatrix}$, matrix $\cT_1$ of sizes $s \times n$ as defined in~\eqref{matrixS}, matrix $\cT_2$ of size $2d \times n$ as defined in Section~\ref{sub:algs:enc}.\\
\textbf{Output:} Set of up to $d$ consecutive positives.

\begin{algorithmic}[1]
\State Set $R = \mathrm{Dec2ConsecutivePositives}_{\mathrm{Bin}}(\bY_1, \cT_1)$. \label{alg:ImprovedUpToD:GetSuperItem} \Comment{Set of super positive items.}
\State If $R$ is empty, return an empty set. Otherwise, the indices of the potential positives are from $(d - 1)\alpha_{\min} + 1$ to $d \alpha_{\max}$, where $\alpha_{\min}$ and $\alpha_{\max}$ are the smallest and the largest indices in $R$. Set $d \alpha_max = n$ if $d \alpha_{\max} > n$. \label{alg:ImprovedUpToD:Potential}
\State For each potential positive $j$, add it to the positive set if $\bY_2(i) = 1$ and $i \equiv j \Mod{(2d)}$, where $i$ is some index in $\{1, 2, \ldots, 2d \}$. \label{alg:ImprovedUpToD:All}
\end{algorithmic}
\end{algorithm}

As in Theorem~\ref{lem:2decoding}, the complexity of decoding $\bY_1$ in Step~\ref{alg:ImprovedUpToD:GetSuperItem} is $O(\log{(n/d)})$. On the other hand, the cost of inferring the smallest and largest indices of the potential positives from $R$ in Step~\ref{alg:ImprovedUpToD:Potential} is just $O(1)$. Therefore, the cost of the first phase is $O(\log{(n/d)})$. Because of the structure of $\cT_2$, each potential positive belongs to only one test and each test contains up to one potential positive. Therefore, by examining $\bY_2$, the true positives among the potential positives will be identified as described in Step~\ref{alg:ImprovedUpToD:All}. It is straightforward that the complexity of the second phase is $O(d)$. In summary, the cost of identify up to $d$ consecutive positives is $O(\log{(n/d)}) + O(d) = O(\log{(n/d)} + d)$.

\section{Algorithms for group testing with the exact number of consecutive positives}
\label{sec:Exact}

In this section, we first present a result on how to identify $d$ positives among $2d$ items then use this result to design a measurement matrix in the second phase of Colbourn's design. The measurement matrix in the first phase can be based on the Gray code as in Section~\ref{sub:revisit:enc} or based on the binary code as in Section~\ref{sub:2Positives:enc1}. The following theorem summarizes the results on finding exactly $d$ consecutive positives among $n$ items.

\begin{theorem}
Let $n$ be a positive integer and $N = \{ 1, 2, \ldots, n \}$ be the set of linearly ordered items with exactly $d$ consecutive positives. Then there exists a deterministic and strongly explicit Gray code-based (binary code-based, respectively) measurement matrix such that the $d$ consecutive positives can be identified with $\ceil{\log_2{\frac{n}{d}}} + d + 3$ ($2 \ceil{\log_2{\frac{n}{d}}} + d$, respectively) tests in $O \left( \log_2{\frac{n}{d}} + d \right)$ time.
\label{thr:2decoding_D}
\end{theorem}

\subsection{Identifying $d$ consecutive positives among $2d$ items}
\label{sub:Exact:locating}

In this section, when the number of (consecutive) positives is known, i.e., $|P| = d$, we show that it only takes $d$ tests to identify the positives among the $2d$ items in time $O(d)$. The number of required tests here is only one larger than the theoretical bound~\cite{colbourn1999group}. The main idea is to allocate each item into a separate test such that the starting positive or the terminal positive can be identified after testing. This idea is summarized below.

\begin{lemma}
Let $N = \{p_1, p_2, \ldots, p_{2d} \}$ be a set of items with the linear order $p_i \prec p_{i + 1}$ for $1 \leq i < d$, so that $N$ has exactly $d$ consecutive positives. Then there exists a deterministic and strongly explicit measurement matrix such that the $d$ consecutive positives can be identified with $d$ tests in $O(d)$ time.
\label{thr:theoreticalBound}
\end{lemma}

\begin{proof}
Let $\cH = (h_{ij})$ be a $d \times 2d$ measurement matrix such that $h_{ii} = 1$ for $i = 1, \ldots, d$ and $\bY = (y_1, y_2, \ldots, y_d)^T$ be the outcome vector obtained from using $\cH$. Matrix $\cH$ can be illustrated in~\eqref{matrixH_odd}. 
\begin{align}
\begin{array}{ccc|ccc}
p_1 & \ldots & p_d & p_{d + 1} & \ldots & p_{2d} \\
\hline
1 & \ & \ & 0 & \ & \ \\
\ & \ddots & \ & \ & \ddots & \ \\
\ & \ & 1 & \ & \ & 0
\end{array} \label{matrixH_odd}
\end{align}

We now show that the matrix $\cH$ in~\eqref{matrixH_odd} can be used to identify to the $d$ consecutive positives. Because of the structure of $\cH$, if the right-most $1$ digit of $\bY^T$ exits, says $i^*$, item $p_{i^*}$ is obviously the index of the starting positive. Otherwise, the starting positive is $p_{d + 1}$. Since we only scan $d$ entries in $\bY$, the time to find the positive set is $O(d)$.
\end{proof}

Instead of using the design in~\eqref{matrixH_odd}, an alternative design of $\cH$ can be used to identify to the $d$ consecutive positives as follows:
\begin{align}
\begin{array}{ccc|ccc}
1 & \ldots & d & d + 1 & \ldots & 2d \\
\hline
0 & \ & \ & 1 & \ & \ \\
\ & \ddots & \ & \ & \ddots & \ \\
\ & \ & 0 & \ & \ & 1
\end{array} \label{matrixH_even}
\end{align}

Indeed, if the right-most $1$ digit of $\bY$ exits, says $i^*$, item $p_{i^*}$ is the index of the terminal positive. Otherwise, the terminal positive is $p_d$. Since we only scan $d$ entries in $\bY$, the time to find the positive set is $O(d)$.

\subsection{Encoding procedure}
\label{sub:Exact:enc}

Similar to the encoding procedure in Section~\ref{sub:algs:enc}, we first create $\kappa = \ceil{n/d}$ super items with linear order in which each super item contains exactly $d$ items, except for the last super item which may contain less than $d$ items. There are up to two super positive items among the $\kappa$ super items.

In the first phase, we can choose the measurement matrix $\cT_1$ as either the one in the first phase of Colbourn's design as in Section~\ref{sub:revisit:enc}, denoted as the Gray code-based design, or the one in the first phase of our design in Section~\ref{sub:algs:enc}, denoted as the binary code-based design. Note that for the Gray code-based design, the number of super items $\kappa$ and the number of items in a super item in Section~\ref{sub:revisit:enc} are replaced with $\ceil{n/d}$ and $d$, respectively. Similar to the arguments in those sections, if $\bY_1$ is the outcome vector of this phase, then $\bY_1 = \cT_1 \odot \bar{\bX}$, where $\bar{\bX} = (x_{\bar{1}}, \ldots, x_{\bar{\kappa}})^T$ and $x_{\bar{j}} = 1$ if and only if the super item $\bar{j}$ is positive, and $x_{\bar{j}} = 0$ otherwise.

In the second phase, a $d \times n$ measurement matrix $\cT_2 = \cH = (h_{ij})$ is created as follows: entry $h_{ij}$ equals to $1$ if and only if $j \equiv i \Mod(2d)$. In other words, each test contains items spaced $2d$ apart in the linear order. The outcome vector by using $\cH$ is $\bY_2 = \cH \odot \bX$.

For the Gray code-based design, the number of required tests for the first phase is $\ceil{\log_2{\ceil{\frac{n}{d}}}} + 3$. Therefore, the total number of tests for two phases in this design is $\ceil{\log_2{\ceil{\frac{n}{d}}}} + d + 3$. With the binary code-based design, the number of required tests for the first phase is $2\ceil{\log_2{\ceil{\frac{n}{d}}}}$. Hence, the total number of tests for two phases in this design is $2\ceil{\log_2{\ceil{\frac{n}{d}}}} + d$.

The final outcome vector by using both $\cT_1$ and $\cT_2$ in the first and second phases is $\bY = \begin{bmatrix} \bY_1 \\ \bY_2 \end{bmatrix}$.

\subsection{Decoding procedure and the decoding complexity}
\label{sub:Exact:dec}

The decoding procedure is described in Algorithm~\ref{alg:DecD}. Step~\ref{alg:DecD:GetSuperItem} is to identify the super positive items among the $\kappa$ super items by decoding $\bY_1$. Depend on the design of the measurement matrix $\cT_1$, one has the corresponding decoding procedure for $\bY_1$. If the number of super positive items is zero, there are no positives in the input set. Therefore, the positive set is empty. Otherwise, the number of super positive items is one or two. This analysis is described in Step~\ref{alg:DecD:empty}. Because the number of consecutive positives is $d$ and each super item contains exactly $d$ consecutive items, Step~\ref{alg:DecD:one} is to return the positive set when the number of super positive items is one. When the number of super positive items is two, i.e., there are $2d$ potential positives, one proceeds to Step~\ref{alg:DecD:two} to identify either the starting positive or the terminal positive. Since each test in $\cT_2$ contains items spaced $2d$ apart in the linear order, the pruned matrix of $\cT_2$ created by taking the corresponding columns of the $2d$ potential positives must have the form either in~\eqref{matrixH_odd} or~\eqref{matrixH_even}. Therefore, by using the decoding procedure in Section~\ref{sub:Exact:locating} which are described in Step~\ref{alg:DecD:two}, the true $d$ positives are identified.

\begin{algorithm}
\caption{Decoding procedure for group testing with exactly $d$ consecutive positives.}
\label{alg:DecD}
\textbf{Input 1:} Outcome vector $\bY = \begin{bmatrix} \bY_1 \\ \bY_2 \end{bmatrix}$, matrix $\cT_1$ is the measurement matrix in the first phase of Colbourn's design defined in Section~\ref{sub:revisit:enc}, matrix $\cT_2$ of size $d \times n$ as defined in Section~\ref{sub:Exact:enc}.\\
\textbf{Input 2:} Outcome vector $\bY = \begin{bmatrix} \bY_1 \\ \bY_2 \end{bmatrix}$, matrix $\cT_1$ is the measurement matrix in the first phase of our design defined in Section~\ref{sub:algs:enc}, matrix $\cT_2$ of size $d \times n$ as defined in Section~\ref{sub:Exact:enc}.\\
\textbf{Output:} Set of $d$ consecutive positives.

\begin{algorithmic}[1]
\State Let $R$ be the set of super positive items obtained by decoding $\bY_1$. If we use Input 1, $R = \mathrm{Dec2ConsecutivePositives}_{\mathrm{Gray}}(\bY_1, \cT_1)$. Otherwise, $R = \mathrm{Dec2ConsecutivePositives}_{\mathrm{Bin}}(\bY_1, \cT_1)$. \label{alg:DecD:GetSuperItem}
\State If $R$ is empty, return an empty set. Otherwise, let $R = \{ \alpha \}$ if $|R| = 1$ and $R = \{ \alpha, \alpha + 1 \}$ if $|R| = 2$. \label{alg:DecD:empty}
\State If $|R| = 1$, return the positive set $\{(\alpha - 1)d + 1, \ldots, \alpha d \}$. \label{alg:DecD:one}
\State If $|R| = 2$, proceed to consider the parity of $\alpha$. When $\alpha$ is odd, the index of the starting positive is $(\alpha - 1)d + i_0$ if the right-most $1$ digit of $\bY_2$ exits, denoted as $i_0$, and is $\alpha d + 1$, otherwise. When $\alpha$ is even, the index of the terminal positive is $\alpha d + i_0$ if the right-most $1$ digit of $\bY_2$ exits, denoted as $i_0$, and is $\alpha d$, otherwise. \label{alg:DecD:two}
\end{algorithmic}
\end{algorithm}

For the Gray code-based design, the complexity of decoding $\bY_1$ is $O(\log{(n/d)})$ as in Theorem~\ref{thr:ColbournX}. For the binary code-based design, the complexity of decoding $\bY_1$ is $O(\log{(n/d)})$ as in Lemma~\ref{lem:2decoding}. On the other hand, the cost of inferring the smallest and largest indices of the potential positives from $R$ in Step~\ref{alg:DecD:empty} is just $O(1)$. Therefore, the cost of the first phase is $O(\log{(n/d)})$ and $O(\log{(n/d)})$, respectively. Moreover, it is obvious that the complexity of the second phase is $O(d)$. Thus, the cost of identifying $d$ consecutive positives by using either the Gray code-based design or the binary code-based design is $O(\log{(n/d)}) + O(d) = O(\log{(n/d)} + d)$.

\section{Simulations}
\label{sec:simul}

We evaluated variations of our proposed schemes and Colbourn's scheme by simulation using $d = 5, 50, 100$, and $N = 2^{16} \approx 66k, 2^{20} \approx 1M, 2^{24} \approx 17M, 2^{28} \approx 300M, 2^{32} \approx 4.3B$, in Matlab R2018a on an Acer Aspire TC-603 desktop PC with a 3.4-GHz Intel Core i7-4770 processor and $8$GB memory. The decoding time was calculated in nanoseconds and averaged over $100$ runs.

\begin{figure}[ht]
\centering
\includegraphics[scale=0.41]{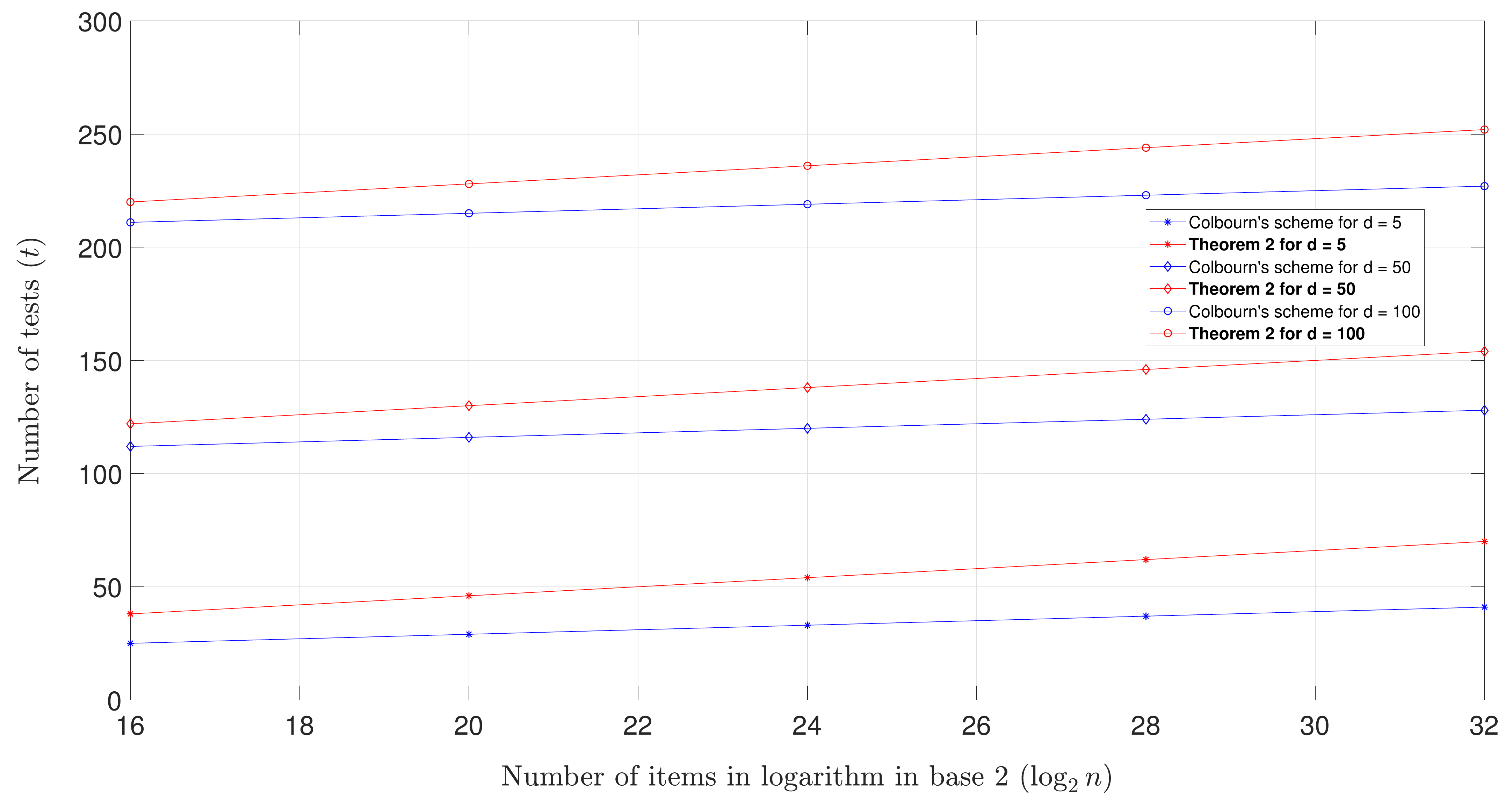}
\caption{The number of tests versus the number of items when the maximum number of consecutive positives is $d$ for $d = 5, 50, 100$.}
\label{fig:UpToD}
\end{figure}

\begin{figure}[ht]
\centering
\includegraphics[scale=0.41]{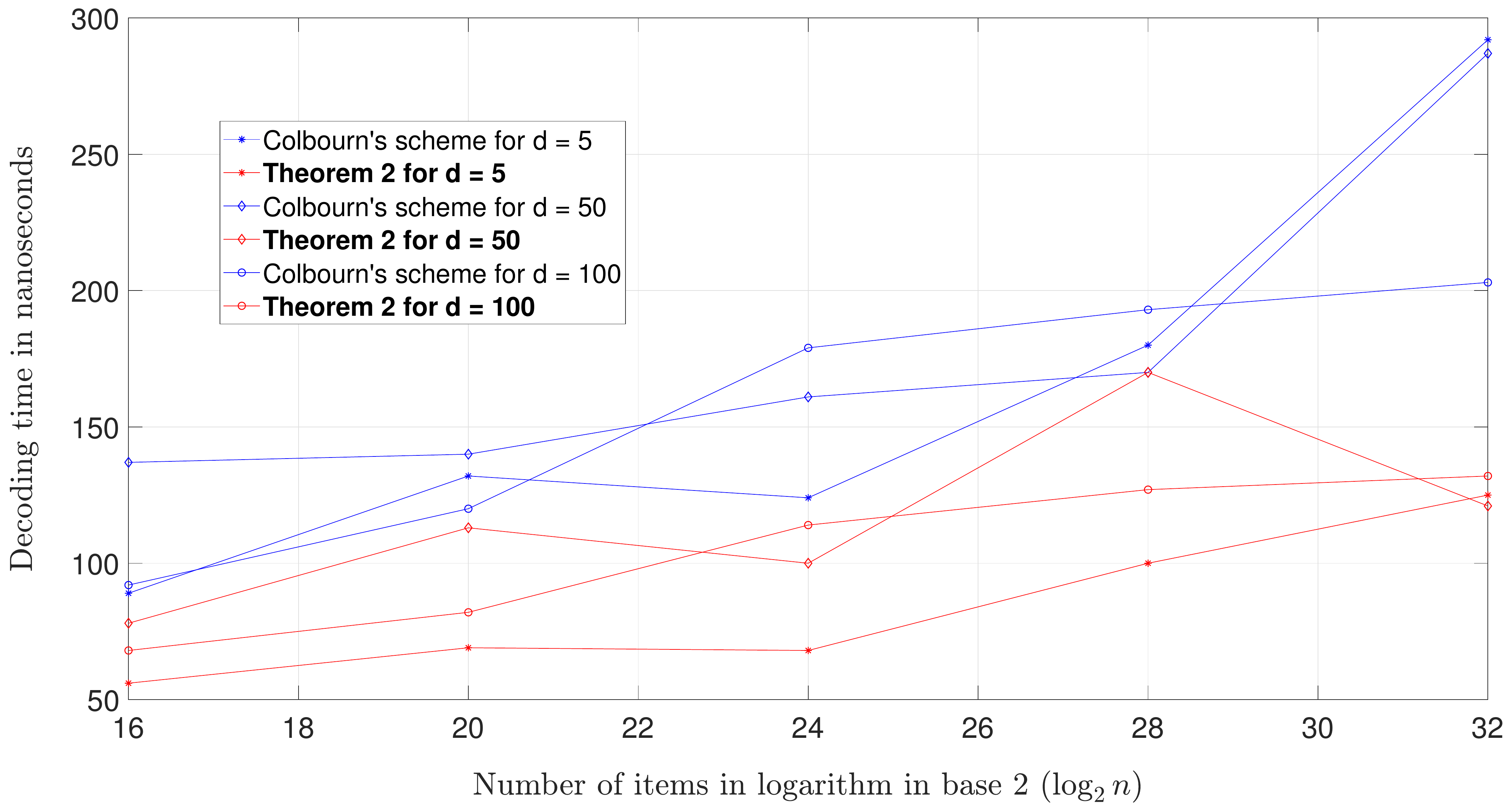}
\caption{Decoding time in nanoseconds versus the number of items when the maximum number of consecutive positives is $d$ for $d = 5, 50, 100$.}
\label{fig:UpToD_Dec}
\end{figure}

When the maximum number of consecutive positives is $d$, the numbers of tests and the decoding times in Theorems~\ref{thr:ColbournX} and~\ref{thr:2decoding_UpToD} are visualized in Fig.~\ref{fig:UpToD} and~\ref{fig:UpToD_Dec}, respectively. While the number of tests in Theorems~\ref{thr:2decoding_UpToD} (our proposed scheme) is slightly larger than in Theorem~\ref{thr:ColbournX}, the decoding time in Theorems~\ref{thr:2decoding_UpToD} is smaller than in Theorem~\ref{thr:ColbournX}. This matches the analysis in Section~\ref{sub:intro:contri} that the decoding procedure in Theorem~\ref{thr:ColbournX} needs to convert the codewords in the outcome vector from the Gray code to the binary code before getting the indices of the positives, whereas our design in Theorems~\ref{thr:2decoding_UpToD} is to directly use the binary code.

\begin{figure}[ht]
\centering
\includegraphics[scale=0.41]{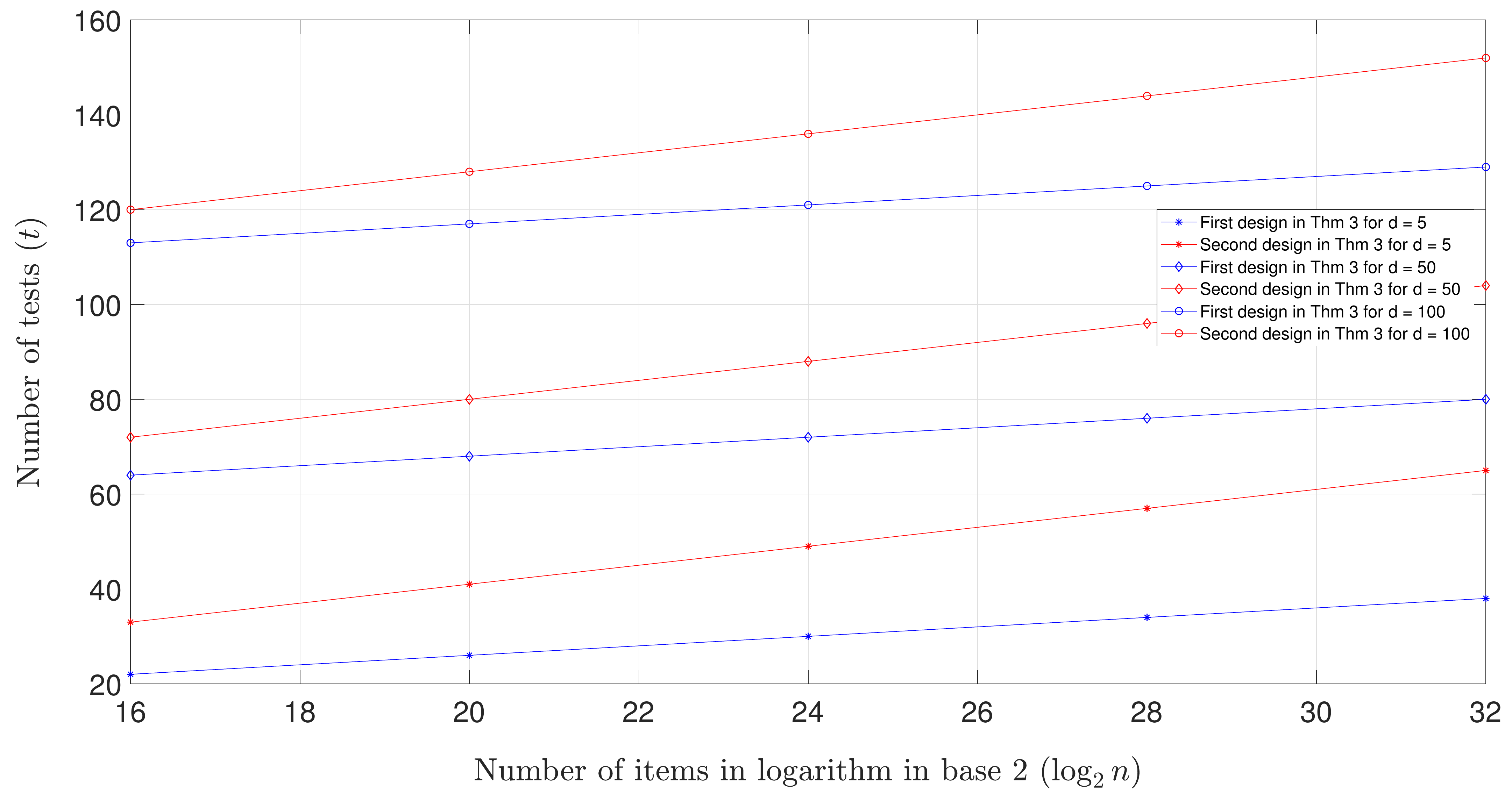}
\caption{The number of tests versus the number of items when the number of consecutive positives is exactly $d$ for $d = 5, 50, 100$.}
\label{fig:exactD}
\end{figure}

\begin{figure}[ht]
\centering
\includegraphics[scale=0.41]{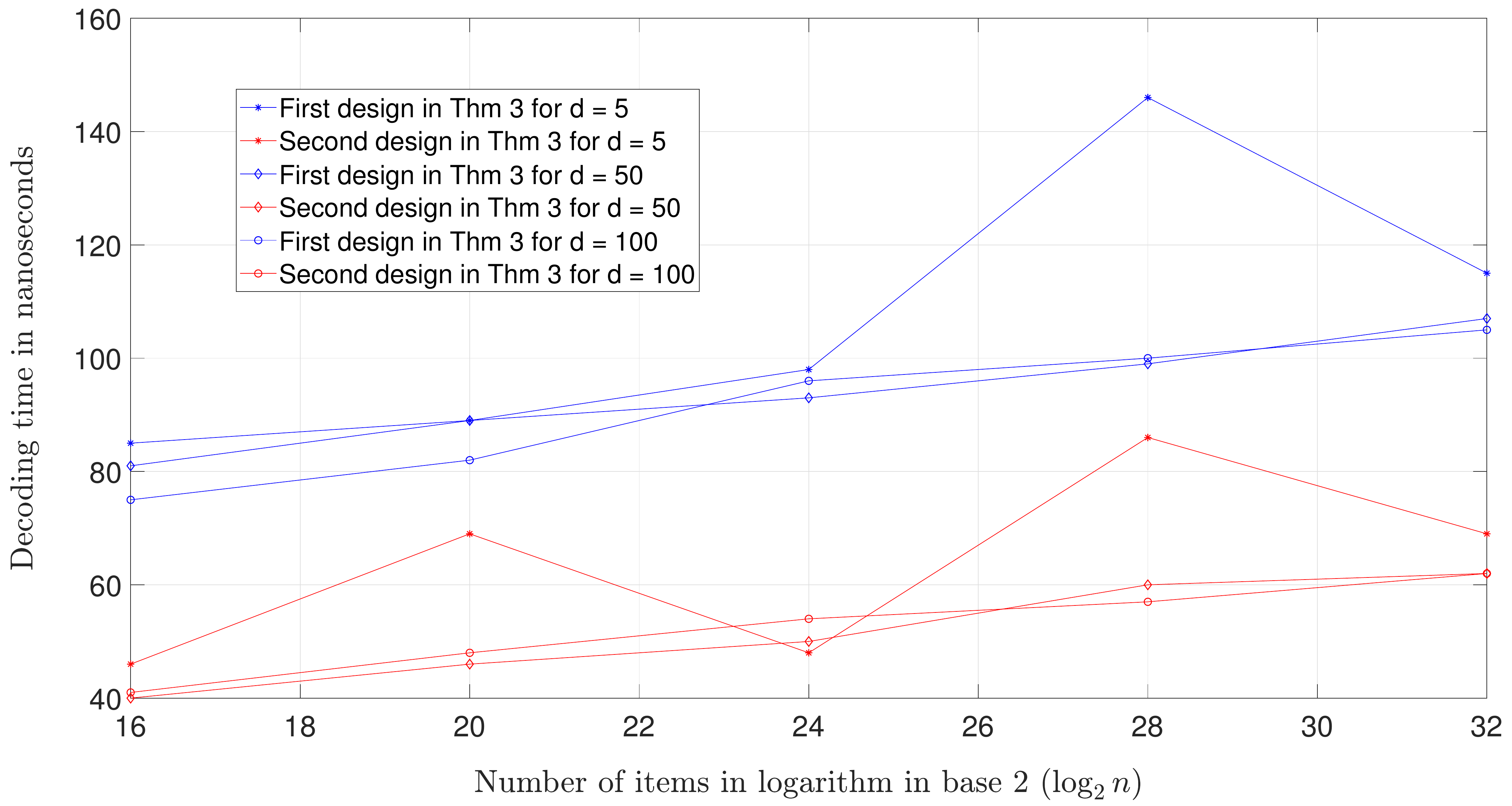}
\caption{Decoding time in nanoseconds versus the number of items when the number of consecutive positives is exactly $d$ for $d = 5, 50, 100$.}
\label{fig:exactD_Dec}
\end{figure}

When the number of consecutive positives is exactly $d$, the numbers of tests and the decoding times of the two designs in Theorem~\ref{thr:2decoding_D} are visualized in Fig.~\ref{fig:exactD} and~\ref{fig:exactD_Dec}, respectively. The term ``Thm'' stands for ``Theorem.'' As shown in Fig.~\ref{fig:exactD}, the number of tests in the Gray code-based design is smaller than in the binary code-based design. However, the decoding time in the Gray code-based design is larger than in the binary code-based design as in Fig.~\ref{fig:exactD_Dec}.

\section{Conclusion}
\label{sec:cls}

In this paper, we have presented various algorithms to efficiently identify consecutive positives when the maximum number of consecutive positives or the number of consecutive positives is known. In the first case, instead of using Gray code, which is a rearrangement of the binary code, we directly use the binary code to construct a measurement matrix to reduce the decoding time in practice. In the second case, the number of tests and the decoding time are almost reduced half compared to the first case. Our simulations verify the efficiency of our proposed designs. An extension of this work to other settings in group testing such as threshold group testing or complex group testing is still an open problem.


\bibliographystyle{ieeetr}
\balance
\bibliography{bibli}

\end{document}